\documentstyle[psfig]{mn2e}

\def\etal{{\it et al. }}

\title[Globular Cluster Kinematics and Dark Matter in NGC 4649]
{The Globular
Cluster Kinematics and Galaxy Dark Matter Content of NGC 4649 (M60)}
\author[Bridges \etal ]{
Terry Bridges$^{1}\thanks{tjb@astro.queensu.ca}$,
Karl Gebhardt$^{2}$,
Ray Sharples$^{3}$,
Favio Raul Faifer$^{4,8}$,
\\ \\ \LARGE
Juan C. Forte$^{4}$,
Michael A. Beasley$^{5}$,
Stephen E. Zepf$^{6}$,
\\ \\ \LARGE
Duncan A. Forbes$^{7}$,
David A. Hanes$^{1}$,
Michael Pierce$^{7}$, \\
\\
$^1$ Department of Physics, Engineering Physics \& Astronomy, Queen's
University, Kingston, ON, K7L 3N6, Canada \\ 
$^2$ Astronomy Department, University of Texas, Austin TX 78712, USA \\ 
$^3$ Department of Physics, University of Durham, South Road,
Durham DH1 3LE, United Kingdom\\ 
$^4$ CONICET and
Facultad de Cs. Astronomicas y Geofisicas, UNLP, Paseo del Bosque
1900, La Plata, Argentina\\ 
$^5$ Instituto de Astrofisica de Canarias, La Laguna 38200, Tenerife,
Spain \\ 
$^6$ Department of Physics and
Astronomy, Michigan State University, East Lansing MI 48824, USA\\ 
$^7$ Centre for Astrophysics \&
Supercomputing, Swinburne University, Hawthorn, VIC 3122, Australia\\ 
$^8$ IALP - CONICET, Argentina \\
}

\begin{document}

\date{}


\maketitle


\begin{abstract}
From observations with the GMOS multi-slit spectrograph on the Gemini
North telescope, we have obtained spectra for 39 globular cluster
candidates in the Virgo giant elliptical galaxy NGC 4649 (M60), of
which 38 are confirmed globular clusters. The clusters extend out to a
radius of 260\arcsec\ (3.5 effective radii). We find no rotation of
the globular cluster system, with an upper limit of v/$\sigma$ $<$ 0.6
at a confidence level of 95\%.
The globular cluster velocity dispersion is constant with radius,
within the uncertainties.  We fit isotropic models to the globular
cluster and stellar kinematics; these models yield a M/L$_V$ around 16 at
200\arcsec\ radius (16 kpc), an increase of a factor of two
from the central M/L. We also use the mass profile as derived
from X-rays to determine the orbital structure. Using axisymmetric
orbit-based models and the X-ray mass profile, we find the orbital
distribution is close to isotropic within 100\arcsec, and becomes
tangentially biased beyond.  Furthermore, when using the X-ray
profile, we find a better fit to the kinematics compared to using
a constant M/L model. Thus, both isotropic and axisymmetric
orbit-based models give support for the presence of a dark matter halo
in NGC 4649.
\end{abstract}


\section{Introduction}
\label{Introduction}

Globular clusters (GCs hereafter) are an excellent way to study the
formation, evolution, and dark matter (DM) content of galaxies of all
types (e.g. Ashman \& Zepf 1998; Harris 2001). They are particularly
well-suited for dynamical studies of early-type galaxies, since they
are found in substantial numbers out to large radii around such
galaxies and are bright enough to be seen at large distances. GCs
complement other dynamical probes such as the stellar light, planetary
nebulae (PNe), and X-ray data (e.g. Gerhard 2005; Romanowsky 2005).
Integrated stellar spectroscopy has the advantage of large 
sample sizes and high
signal-to-noise, which allow the derivation of higher-order moments of
the velocity distribution, but the faintness of the stellar halo
restricts measurement to within a few effective radii of galaxy
centers (e.g. Kronawitter et al. 2000; Gerhard et al. 2001). X-ray
data can probe further out, and recent high-quality data from ROSAT,
ASCA, and {\it Chandra} have shown that extended dark halos are a common
feature of all luminous ellipticals (Mathews \& Brighenti
2003; Humphrey et al. 2006). 
However, accurate X-ray mass determinations require
high-quality gas density and temperature profiles, along with the
assumption of hydrostatic equilibrium. As well, only massive
ellipticals have hot diffuse gas halos for mass estimates.

PNe also have extended spatial distributions, and are easily
identified because of their strong emission in the [OIII] 4959/5007
lines. 
Romanowsky et al. (2003), using the Planetary Nebula Spectrograph
(Douglas et al. 2002), found 
that three intermediate-luminosity
ellipticals (NGCs 821, 3379, and 4494) have
declining PNe velocity dispersion profiles, consistent with
simple models with little or no DM.  This result is in
conflict with the conventional model of galaxy formation, whereby
galaxies form in dark matter halos. However, numerical simulations
by Dekel et al. (2005) show that spiral-spiral mergers may lead to
halo stars having highly radial orbits, as they are tidally ejected
from the inner regions during the merger.
Observed PNe in ellipticals, as part of this stellar halo, would then
have line-of-sight velocity dispersions lower than that of the
underlying DM. Older tracers such as GCs
are expected to have higher velocity dispersions,
more representative of the DM. In support of this, 
Sambhus, Gerhard \& Mendez (2006) have recently found
that the bright and faint PNe in NGC 4697 have different kinematics,
and suggest that the bright PNe are a younger population perhaps formed
in tidal structures.
Our group (Pierce et al. 2006a) 
has found that the GC velocity dispersion profile in NGC 3379
increases with radius, and non-parametric, isotropic models yield a 
significant increase in the M/L ratio at large radii. 
A similar result for the NGC 3379 GCs
has also been found by Bergond et al.(2006), from FLAMES/VLT data.
Note also that Peng, Ford \& Freeman (2004a) found a relatively low
M/L ratio of 10$-$15 for NGC 5128 from PNe velocities. 
It is thus very important 
to compare GC and PNe kinematics in a larger number of galaxies, to gain 
a more complete understanding of host galaxy dynamics.

GC velocities have been used to study the dynamics of several
early-type galaxies, including M87 (Cohen 2000; Cote et al. 2001), 
NGC 4472 (Sharples et al. 1998; Zepf et al. 2000; Cote et al. 2003),
NGC 1399 (Richtler et al. 2004), NGC 5128 (Peng, Ford \& Freeman 2004b), and 
NGC 3379 (Pierce et al. 2006a; Bergond et al. 2006); 
see Zepf (2003) for a review. 
From these studies it is consistently found that the GC velocity
dispersions are either constant or increasing with radius. This
is strong evidence for DM halos in these galaxies,
supporting X-ray data. Merger simulations
(Hernquist \& Bolte 1993; Vitvitska et al. 2002; Bekki et al. 2005) 
predict that ellipticals should have significant
amounts of angular momentum at large radius. 
However, rotation is seen in some GC systems (e.g. NGC 5128;
M87), but not in others (e.g. NGC 1399; NGC 3379; NGC 4472). 
In some cases, the lack of detected rotation may be due to small
numbers of GCs; but NGC 4472 and NGC 1399 have hundreds of GC velocities,
allowing tight constraints on rotation except at the outer reaches of the
datasets.
The situation is even more complex when the kinematics
of metal-poor (MP) and metal-rich (MR) GCs are studied separately. In most GC
systems, the MP and MR GCs are different in terms of their rotation
and velocity dispersions, and these differences often depend on
galactocentric radius. Usually the MP GCs are dynamically hotter than
the MR GCs, as might be expected given their generally shallower
spatial profiles (e.g. NGC 5128, NGC 1399, NGC 4472). In some cases
(e.g. NGC 4472, M87) the blue and red clusters may be counter-rotating
at some radii. In only a few galaxies with the largest GC samples can
we say anything about GC orbital anisotropy; in NGC 1399, NGC 4472,
and M87 the GCs are consistent with isotropic orbits (though there are
hints that the orbital behavior may be different for MP and MR
sub-samples).

This paper continues our Gemini/GMOS study of the GC system of NGC
4649 (M60), a bright elliptical located in a subcluster in the eastern
part of the Virgo cluster (see Forbes et al. 2004 for further details
of NGC 4649 and photometric studies of its GC system). NGC 4649 is
well-studied in X-rays, with ROSAT (Trinchieri, Fabbiano \& Kim 1997;
Di Matteo \& Fabian 1997; Bohringer et al. 2000; O'Sullivan, Forbes,
\& Ponman 2001) and {\it Chandra} (Irwin, Athey \& Bregman 2003; 
Sarazin et al. 2003;
Swartz et al. 2004; Randall, Sarazin \& Irwin 2004; 
Gutierrez \& Lopez-Corredoira
2005; Humphrey et al. 2006) data. Trinchieri, Fabbiano \& Kim (1997),
Bohringer et al. (2000), and Humphrey et al. (2006) all find 
evidence for an extended dark matter halo
in NGC 4649; Humphrey et al. (2006) find an enclosed mass of 
~1 $\times$ 10$^{12}$ M$_\odot$ and a M/L $\sim$ 5 in the K band
at 260\arcsec\ (20 kpc).
{\it Chandra} data have also revealed over 100 low mass x-ray binaries
(LMXRBs) in NGC 4649 (Sarazin et al. 2003; Randall, Sarazin \& Irwin 2004), 
and we will use our larger GMOS GC imaging database (Forbes et al. 2004)
to match with the {\it Chandra} LMXRBs in a future paper 
(Bridges et al. 2006, in preparation).

Previous studies of the stellar kinematics in NGC
4649 have found major axis stellar rotation of 
100$-$120 km/sec out to 45$-$100\arcsec, a high central velocity
dispersion of 350$-$400 km/sec, and a declining velocity dispersion
profile (Fisher, Illingworth \& Franx 1995; de Bruyne et al. 2001; Pinkney et
al. 2003); de Bruyne et al. also find modest minor axis rotation
of $\sim$ 30 km/sec. de Bruyne et al. use 3-integral
axisymmetric models to find that a dark matter halo gives the best fit
to the kinematic data, though a constant M/L ratio model is also
consistent with the data. We will use the data from Pinkney et al. to
combine with our GC velocities and X-ray data (Humphrey et al. 2006)
to study the dynamics of NGC 4649 out
to large radius (Section \ref{mlratio}).

Forbes et al. (2004) presented GMOS photometry for $\sim$ 1000 GC
candidates, and confirmed previous findings of bimodality in the GC
colour distribution.  Forbes et al. also found that the red GCs are
more concentrated to the galaxy centre than the blue GCs, as is found
in many other galaxies. The underlying galaxy light has a similar density
profile slope as the red GCs.
These GMOS images were used to select GC
candidates for follow-up spectroscopy using GMOS in multi-slit mode
(see Section~\ref{selection}). This paper presents spectra for 38
confirmed NGC 4649 GCs, and discusses the GC kinematics (rotation and
velocity dispersion profiles; Section~\ref{kinematics}) and dark
matter content of NGC 4649 (Section~\ref{mlratio}). In Section
~\ref{discussion} we discuss our main results in more detail, while
Section~\ref{conclusions} presents our conclusions. A companion
paper (Pierce et al. 2006b) uses spectroscopic line
indices to address the ages, abundances, and abundance ratios of the
confirmed GCs. We adopt an SBF
distance of D = $17.29^{+0.58}_{-0.56}$ Mpc for NGC 4649 (Mei et al.
2006, in preparation), taken from the ACS Virgo Cluster Survey
(Cote et al. 2004); this gives an image scale of 11.9\arcsec/kpc.
This is slightly larger than the distance of 16.8 Mpc we used in our 
photometric study of the NGC 4649 GCs (Forbes et al. 2004). 

\section{Multi-Object Spectroscopy}

\subsection{Object Selection}
\label{selection}

\begin{figure*}
\centerline{\psfig{figure=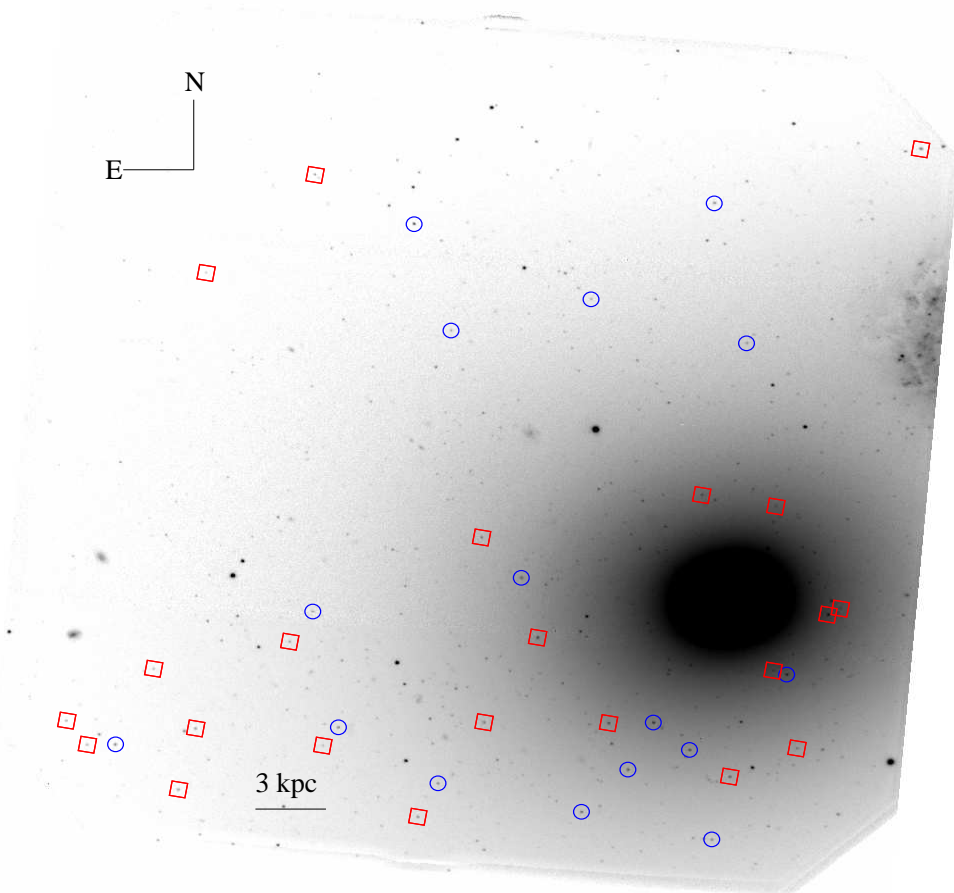,width=15cm}}
\caption{NGC 4649 Field 1. Field orientation is shown at the 
top left of the figure. For comparison with Table \ref{cattable},
X (Column 3) increases from bottom to top, while Y (Column 4) increases
from left to right.
NGC 4647 can be seen at the right-hand edge of the frame to the NW
of NGC 4649. Confirmed blue globular clusters are indicated by (blue) circles, 
and red
globular clusters by (red) squares; the split between blue and red is at 
g-i = 1.0.
}
\label{field-image}
\end{figure*}

GC candidates for follow-up GMOS multi-slit spectroscopy were selected
from our GMOS images taken on 2002 April 10, 11, and 14 (Gemini
program reference number GN-2002A-Q-13). These consist of 4$\times$
120 sec exposures in Sloan g and i filters for three
fields. Each field is 5.5 $\times$ 5.5\arcmin\ and the pixel
scale is 0.07\arcsec/pixel; see Forbes et
al. (2004) for more details about the preimaging.  To date, we have
only obtained spectroscopy for Field 1, centered 1.9\arcmin\ to the NE
of the galaxy center (see Figure~\ref{field-image}); the seeing for
Field 1 was $\sim$ 0.6\arcsec. We used the co-added GMOS images
provided by Gemini staff for our object selection.

Here is a brief discussion of the procedure we have used to identify a
clean sample of GC candidates for spectroscopy. We first
median-filtered each image using a square filter $\sim$ 10 times the
image FWHM, and subtracted the medianed image from the original. We
then used DAOPHOT to obtain photometry for point-like objects
(DAOFIND, PHOT, and ALLSTAR). Aperture corrections were then obtained
to obtain total magnitudes.  Our photometry was calibrated onto F555W
(V) and F814W (I), using HST WFPC2 photometry kindly provided by
Soeren Larsen (Larsen et al. 2001). This calibration was first done on
our Field 2, which has the most overlap with the HST data; we then
used overlapping regions between Field 1 and Field 2 to calibrate
Field 1. We next matched the g (V) and i (I) images.
The final result is photometry for $\sim$ 800 point-like sources in
Field 1, with V, I, and (V$-$I) colours good to $\pm$ 0.1 mag
(sufficient for purposes of object selection).

From this input catalogue, we obtained a final sample of GC candidates
by magnitude and colour selection. We selected a magnitude range from
19 $<$ V $<$ 22.5; the bright cutoff corresponds to where we expect
GCs to start appearing at the distance of NGC 4649 (for a distance of
17.3 Mpc, V=19 is $\sim$ 3.3$\sigma$ from the turnover in the GC
luminosity function),
while the faint cutoff is chosen to give at least a S/N of 5 in an 8
hour exposure. Our colour range of 0.75 $<$ V$-$I $<$ 1.4 was chosen
to include all Galactic GCs (Harris 1996), with
some allowance for more metal-poor (bluer) or more metal-rich (redder)
GCs. This color range also encompasses the vast majority of GCs seen
in other elliptical galaxies (e.g. Kundu \& Whitmore 2001; Larsen et
al. 2001). Finally, obviously extended sources were eliminated by
eye. After these cuts, 250 objects remained as NGC 4649 GC candidates. 

A GMOS multi-slit mask was made for Field 1 using the standard Gemini
GMMPS software. In addition to our GC candidates, we included three
stars with 15.5 $<$ V $<$ 18.5 for mask
alignment. Using the B600 grism with a central wavelength of 5000 \AA,
a slit width of 1\arcsec, and a minimum slit length of 4\arcsec, we
were able to allocate 39 slits in Field 1 and the three alignment stars.

\subsection{Spectroscopic Observations and Data Reduction}

GMOS spectroscopy was obtained for Field 1 on 2003 May 31, 2003 June 1, 
2003 June 4, and 2003 June 27, as part of program GN-2003A-Q-22. A total of 8
$\times$ 1800 sec exposure was obtained with a central wavelength of
5000 \AA, and 8 $\times$ 1800 sec at 5050 \AA; we thus have a total of 8
hours on-source time. We used the B600\_G5303 grism giving a dispersion
of 0.45 \AA/pixel and a resolution of $\sim$ 5.5
\AA. Spectra typically cover a typical range from 3300$-$5900
\AA, but the wavelength range depends on the slit position on
the mask and some spectra reach up to $\sim$ 7000 \AA.  The
seeing ranged from 0.65 to 0.9\arcsec\ over the 4 nights. Bias frames,
dome flatfields and Copper-Argon (CuAr) arcs were also taken for
calibration. Typical wavelength residuals of 0.1 \AA~ were
achieved.

Data reduction was done in the following way, using the gemini/gmos
IRAF package (Version 1.6):

\begin{itemize}

\item GSFLAT was used to make a normalized spectroscopic flatfield
from individual flatfield exposures (typically 3 or 4)

\item Object frames were bias-subtracted and flatfielded using
GSREDUCE

\item The arc frame was bias-subtracted, but not flatfielded, using
GSREDUCE

\item GSWAVELENGTH was used to establish the wavelength calibration
from the Cu-Ar arc frames. This is a 2-D fit, and several iterations
are required per aperture. This is the most tedious and interactive
part of the reduction.

\item The object frames were then wavelength calibrated using
GSTRANSFORM

\item The object frames were then sky-subtracted interactively, using
GSSKYSUB

\item GEMCOMBINE was used to create a combined exposure from
individual object frames, to act as a reference for extraction of the
object spectra (after checking to make sure there are no shifts
between the object frames)

\item GSEXTRACT was used to extract 1-D spectra from each object
frame, using the reference frame created in the previous step; optimal
(variance) extraction was found to give the best results.

\item Finally, SCOMBINE was used to combine spectra from the different
object frames; we did a median combine, used avsigclip for bad pixel
rejection, and scaled each individual spectrum by the median of the
counts between 4500$-$5000 \AA.

\end{itemize}

The final spectra have
S/N ranging from 5-21 per \AA~ at 5000 \AA, with a median
S/N of 11.5 per \AA.

\subsection{GC Velocity Determinations}
\label{veldeterm}

To measure the GC velocities, we used the FXCOR task in IRAF. Due to
prohibitive time overheads, we did not observe
radial velocity standard stars with GMOS. Therefore, we
used six Bruzual \& Charlot (2003) model spectra for cross-correlation
templates. These model spectra have metallicities [Fe/H] of $-$1.64,
$-$0.33, and +0.1, for ages of 5 and 14 Gyr. We used the average
of these 6 templates for our final velocity (taking a weighted average
only changes the velocities by $\sim$ 1 km/sec); all cross-correlation
peak heights were higher than 0.1 and are therefore very secure. The
velocity errors were taken to be the average of the errors returned 
by FXCOR for the 6 templates, and thus only reflect the statistical
uncertainties in these fits.
The final velocities with their errors are given in the last
two columns of Table~\ref{cattable}.

There was only one background object (a QSO at z $\sim$ 0.5) among the 39
spectra. Our low contamination rate of 2.5\% is attributable to 
the good seeing of our GMOS images, and careful
colour and magnitude selection. The remaining 38 objects have 
heliocentric velocities ranging from 484 to 1511 km/sec, and as we
discuss in Section ~\ref{kinematics}, we believe that these are
all GCs belonging to NGC 4649.

Table~\ref{cattable} presents positions, photometry, and
heliocentric velocities for our 38 confirmed NGC 4649 GCs. Note
that the photometry in Table~\ref{cattable} is GMOS Sloan
g and g$-$i, taken from Forbes
et al. (2004); errors in g and g$-$i
are less than 0.05 mag. 
An approximate conversion to HST/WFPC2 V (F555W) and 
I (F814W) is given by: V = g $-$ 0.75 $\pm$ 0.1, and
V$-$I =  g$-$i $-$ 0.02 $\pm$ 0.1.
Figure~\ref{field-image} shows the spatial distribution of the confirmed
GCs.

Two objects deserves separate mention. Object \#558 is marginally
resolved on our GMOS images, and clearly extended on HST/ACS F475W and
F850LP images (Proposal 9401: note that the HST/ACS images only became
public in July 2004, and we did not have access to them for our
spectroscopic object selection). Its velocity (1088 km/sec) is
consistent with both that of NGC 4649 (1117 km/sec) or the Virgo
cluster (1079 km/sec). Its apparent magnitude is V = 21.0, giving it
an absolute magnitude M$_V$ = $-$10.1 for a Virgo distance of 16.8
Mpc. It is our bluest object, with g-i = 0.74. Thus, its magnitude and
colour are consistent with either a luminous GC or a faint (stripped?)
dwarf galaxy. Stellar population model fits give an old age and low
metallicity for \#558, but the fits are uncertain (Pierce et
al. 2006b).  For the moment, we retain this object in our kinematic
sample.  Object \#1574 is in fact closer in both position and velocity
to NGC 4647 than to NGC 4649, and could plausibly be a GC associated
with NGC 4647. Exclusion of \#1574 changes the mean velocity and
velocity dispersions by less than 10 km/s, and makes no difference to
the dynamical modelling in Section \ref{mlratio}; thus, we retain 
this object also in our kinematic sample.

\begin{table*}
\centering
\begin{minipage}{140mm}
\caption{Confirmed globular clusters around NGC 4649. Cluster ID,
RA, Dec, X, Y, g, g$ - $i, and heliocentric
velocities and errors are presented for 38 confirmed GCs. The
g and g - i photometry is from Forbes
et al. (2004), while the heliocentric velocities are from this work.
To convert to HST/WFPC2 V (F555W) and
I (F814W) use: V = g $-$ 0.75 $\pm$ 0.1, and
V$-$I =  g$-$i $-$ 0.02 $\pm$ 0.1.
The X and Y positions refer to the image shown in
Figure~\ref{field-image}; the galaxy center is approximately at
(X,Y) = (2405,3664). The RA and Dec have been calculated using 
$\sim$ 15 USNO B1.0 stars found in the frame, and the positions 
are good to $\sim$ 1\arcsec.}
\label{cattable}
\begin{tabular}{ccccccccc}
\hline
ID & R.A. & Dec. & X & Y & g & g$-$i &
V$_{helio}$ & V$_{err}$ \\
  & (J2000) & (J2000) & (pix) & (pix) & (mag) & (mag) & (km/sec) & (km/sec) \\
\hline
    89 & 190.983978 &  11.536740 & 1324.96 &  478.67 &  22.72 &  1.08 & 1199.2 &   43.4 \\
   124 & 190.981033 &  11.536771 & 1339.40 &  621.02 &  21.91 &  0.88 &  833.4 &   59.6 \\
    68 & 190.986160 &  11.539418 & 1447.41 &  362.65 &  22.58 &  1.22 &  649.3 &   33.6 \\
   148 & 190.977036 &  11.545153 & 1769.64 &  777.98 &  22.74 &  1.14 & 1184.8 &   46.8 \\
   175 & 190.974426 &  11.531744 & 1120.73 &  961.31 &  22.29 &  1.18 &  852.3 &   37.6 \\
   183 & 190.972626 &  11.538513 & 1461.97 & 1019.24 &  22.26 &  1.13 & 1299.4 &   35.8 \\
   158 & 190.971558 &  11.589099 & 3957.80 &  853.95 &  23.21 &  1.16 & 1261.0 &   46.0 \\
   360 & 190.949738 &  11.594447 & 4316.58 & 1883.79 &  21.20 &  0.91 & 1270.1 &   40.1 \\
   329 & 190.957642 &  11.538593 & 1531.35 & 1741.23 &  21.93 &  0.88 & 1511.2 &   65.3 \\
   277 & 190.962753 &  11.548118 & 1978.06 & 1454.03 &  22.59 &  1.04 & 1299.1 &   46.5 \\
   251 & 190.960129 &  11.599953 & 4542.24 & 1358.58 &  22.29 &  1.09 & 1099.7 &   43.9 \\
   298 & 190.960342 &  11.551459 & 2153.17 & 1556.03 &  22.58 &  0.90 & 1063.0 &   52.5 \\
   318 & 190.959274 &  11.536565 & 1424.30 & 1671.14 &  23.08 &  1.00 & 1173.4 &   48.6 \\
   606 & 190.936737 &  11.548500 & 2110.58 & 2706.89 &  21.36 &  1.18 &  627.0 &   39.6 \\
   558 & 190.938461 &  11.555140 & 2430.02 & 2595.52 &  21.75 &  0.74 & 1087.8 &   50.8 \\
   434 & 190.949310 &  11.528630 & 1077.12 & 2185.82 &  22.18 &  1.04 &  925.9 &   36.6 \\
   462 & 190.947189 &  11.532343 & 1269.20 & 2272.22 &  22.18 &  0.92 & 1112.8 &   45.4 \\
   517 & 190.942352 &  11.539107 & 1623.47 & 2476.30 &  22.03 &  1.14 &  857.2 &   56.3 \\
   412 & 190.945847 &  11.582625 & 3751.32 & 2121.39 &  22.48 &  0.95 &  483.6 &   47.5 \\
   502 & 190.942627 &  11.559628 & 2632.89 & 2375.35 &  22.10 &  1.17 &  688.6 &   43.5 \\
   640 & 190.931168 &  11.586047 & 3984.06 & 2815.36 &  22.61 &  0.79 & 1015.5 &  104.5 \\
   740 & 190.932129 &  11.529132 & 1176.86 & 3012.11 &  21.68 &  0.98 &  962.1 &   44.8 \\
   806 & 190.929291 &  11.538971 & 1673.81 & 3107.00 &  21.50 &  1.18 & 1197.4 &   46.2 \\
   899 & 190.927261 &  11.533805 & 1428.31 & 3227.10 &  21.44 &  0.92 & 1372.0 &   42.7 \\
   975 & 190.924591 &  11.539021 & 1696.82 & 3333.25 &  20.99 &  0.92 & 1052.2 &   42.3 \\
  1063 & 190.920807 &  11.535969 & 1563.10 & 3529.46 &  21.37 &  0.94 &  953.9 &   47.3 \\
  1011 & 190.918243 &  11.596663 & 4563.36 & 3393.26 &  21.95 &  0.94 & 1135.0 &   45.7 \\
  1037 & 190.919540 &  11.564267 & 2962.26 & 3469.39 &  21.97 &  1.13 &  667.9 &   63.4 \\
  1145 & 190.916580 &  11.532989 & 1434.77 & 3745.73 &  21.60 &  1.08 & 1017.1 &   42.6 \\
  1252 & 190.910614 &  11.544307 & 2018.25 & 3985.40 &  21.27 &  0.89 &  926.8 &   47.3 \\
  1384 & 190.906281 &  11.550952 & 2364.45 & 4165.70 &  20.91 &  1.03 & 1134.2 &   47.0 \\
  1126 & 190.918457 &  11.526026 & 1083.67 & 3685.07 &  22.21 &  0.90 & 1122.0 &   58.2 \\
  1298 & 190.909515 &  11.536103 & 1619.00 & 4073.00 &  22.25 &  1.13 & 1346.1 &   47.9 \\
  1211 & 190.912033 &  11.544750 & 2033.87 & 3914.83 &  22.29 &  1.01 & 1324.5 &   38.0 \\
  1098 & 190.914841 &  11.581110 & 3812.20 & 3623.63 &  22.27 &  0.77 &  889.3 &   69.3 \\
  1182 & 190.911743 &  11.562973 & 2932.52 & 3850.43 &  22.32 &  1.15 & 1305.2 &   39.3 \\
  1574 & 190.896591 &  11.602605 & 4950.56 & 4411.99 &  21.65 &  1.01 & 1454.0 &   45.1 \\
  1443 & 190.904984 &  11.551611 & 2402.56 & 4225.51 &  22.65 &  1.28 &  703.0 &   53.9 \\
\hline
\end{tabular}
\end{minipage}
\end{table*}

\section{Results}

\subsection{Kinematic Analysis}
\label{kinematics}

\subsubsection{Velocity Distribution and Galaxy Membership}

\begin{figure*}
\centerline{\psfig{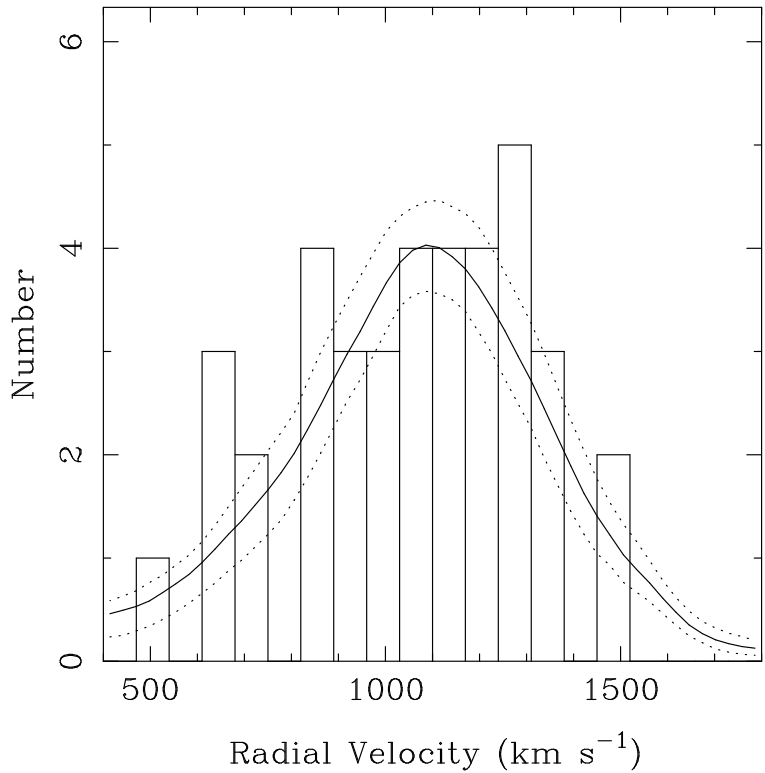}}
\caption{Velocity distribution of NGC 4649 globular clusters. The bars
are a conventional histogram, while the solid and dotted lines are a 
generalized
velocity histogram with 68\% confidence intervals.}
\label{vel_dist}
\end{figure*}

Figure~\ref{vel_dist} shows both a conventional histogram and a generalized
histogram of the velocities for the 38 NGC 4649
GC candidates. The generalized histogram uses an adaptive kernel technique
to provide a non-parametric density estimate on a grid; it is explained
in detail in Gebhardt et al. (1996). The 68\% uncertainties result from
Monte Carlo simulations. The systemic velocity of 
NGC 4649 is 1117 $\pm$ 6 km/s (Trager et al. 2000), the biweight
mean velocity for all 38 GC candidates is 1066 $\pm$ 45 km/s, and
the median GC candidate velocity is 1094 km/s. 

We first need to determine which of our 38 GC candidates are
indeed bona-fide NGC 4649 GCs. We first note that the 
generalized histogram in Figure~\ref{vel_dist}
shows a very smooth distribution, with no obvious groups of low or
high velocity non-clusters. Second, our lowest candidate velocity is
483 km/s, and it is very unlikely that a Milky Way field star would
have this high a velocity. For instance, 99\% of SDSS stars have 
velocities less than 483 km/s. We have also ran Besancon models
(Robin et al. 2003) for the velocity distribution of halo stars towards
NGC 4649 and none had a velocity as high as 483 km/sec. Finally, 
none of the 41,000+ stars in the RAVE survey have a velocity as 
high as 483 km/s (M. Williams 2005, private communication).

Another approach is to try KMM mixture modelling (McLachlan \& Basford
1988; Ashman, Bird \& Zepf 1994), to see if there is any evidence for more than
one velocity group. 
We ran the KMM code for two
groups, with the first group consisting of the 6 candidates with
velocities 484 $\leq$ 703 km/sec, trying both homoscedastic
(common covariance) and heteroscedastic groups (independent covariance).
We also tried running KMM with the first group consisting solely of the
lowest velocity candidate at 483 km/sec, but the result was identical
to the first case. 
In no case did we find that two groups were a statistically better
fit to the velocity distribution than one (P-values of 0.25 and 0.52
for homoscedastic and heteroscedastic groups respectively). 

We conclude that all of the 38 GC candidates are indeed bona-fide
GCs belonging to NGC 4649 (but note that there is uncertainty about
whether object \#1574 belongs to NGC 4649 or NGC 4647, as discussed
in Section \ref{veldeterm}).

\subsubsection{Velocity Dispersion}
\label{dispersion}

We have used the ROSTAT code (Beers, Flynn \& Gebhardt 1990) to
calculate the biweight velocity dispersion for the NGC 4649 GCs. We
have done this for all of the GCs, and for the blue and red GCs
separately. The division
between blue and red GCs is at (g-i) = 1.0, based on the bimodality in
the GC colour distribution seen in Forbes et al. (2004). The velocity
dispersion for all 38 GCs (256 $\pm$ 29 km/sec) is comparable to the
dispersions found in other early-type galaxies. The
velocity dispersion for the red GCs (288 $\pm$ 43 km/s) is 
higher than for the blue GCs (205 $\pm$ 65 km/s), in contrast to
other luminous ellipticals.
To see how sensitive these results are to our colour boundary, we have
set the boundary at (g-i) = 0.95 and (g-i) = 1.05; the blue/red
dispersions are 181 $\pm$ 50/297 $\pm$ 39, and 222 $\pm$ 44/286 $\pm$
53, for a boundary of (g-i) = 0.95/1.05. The dispersions, especially
for the blue GCs, are sensitive to the adopted colour boundary, but
the difference in dispersion between the red and blue GCs persists
for all 3 choices.  The red GCs thus have a higher
velocity dispersion than the blue GCs of 64$-$116 km/sec, but at a
low significance level ranging from 0.9 to 1.8$\sigma$. 

\begin{figure*}
\centerline{\psfig{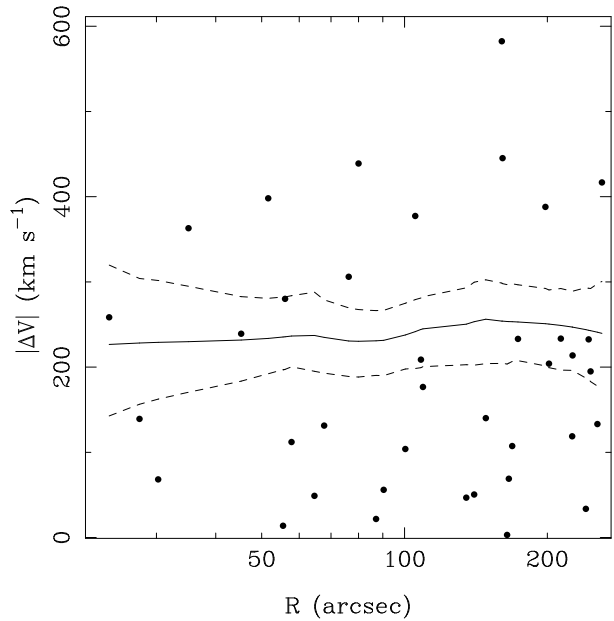}}
\caption{
Velocity dispersion vs radius for NGC 4649 globular clusters. 
The points are the GC velocities, and the solid curve and dashed lines
are the lowess estimator fit with 68\% confidence bands.
}
\label{dis_profile}
\end{figure*}

Figure~\ref{dis_profile} shows the GC velocity dispersion radial
profiles.
This dispersion
profile was derived using a lowess estimator, which runs a radial
window function through the data to estimate the velocity squared (see
Gebhardt \& Fischer 1995, and Pierce et al. 2006a for further
details).  Figure~\ref{dis_profile} shows that the GC velocity
dispersion profile is constant with radius within the
errors. We have checked whether removal of
the young and intermediate age clusters, as discussed in Pierce et
al. (2006b), changes the dispersion values; within the uncertainties
there is no change in the results.

\subsubsection{Rotation of the Globular Cluster System}

We have searched for rotation in the GC system
by doing non-linear least squares fits to the following equation:

$$V(\theta) = V_{rot}sin(\theta - \theta_0) + V_0$$

\noindent where V$_{rot}$ is the GC rotation velocity (amplitude), 
$\theta$ is the
GC azimuthal angle, $\theta_0$ is
the position angle of the line of nodes, and V$_0$ is the systemic
velocity of NGC 4649. This determines the
best-fitting flat rotation curve; see Zepf et al. (2000) for more details. 

\begin{figure*}
\centerline{\psfig{figure=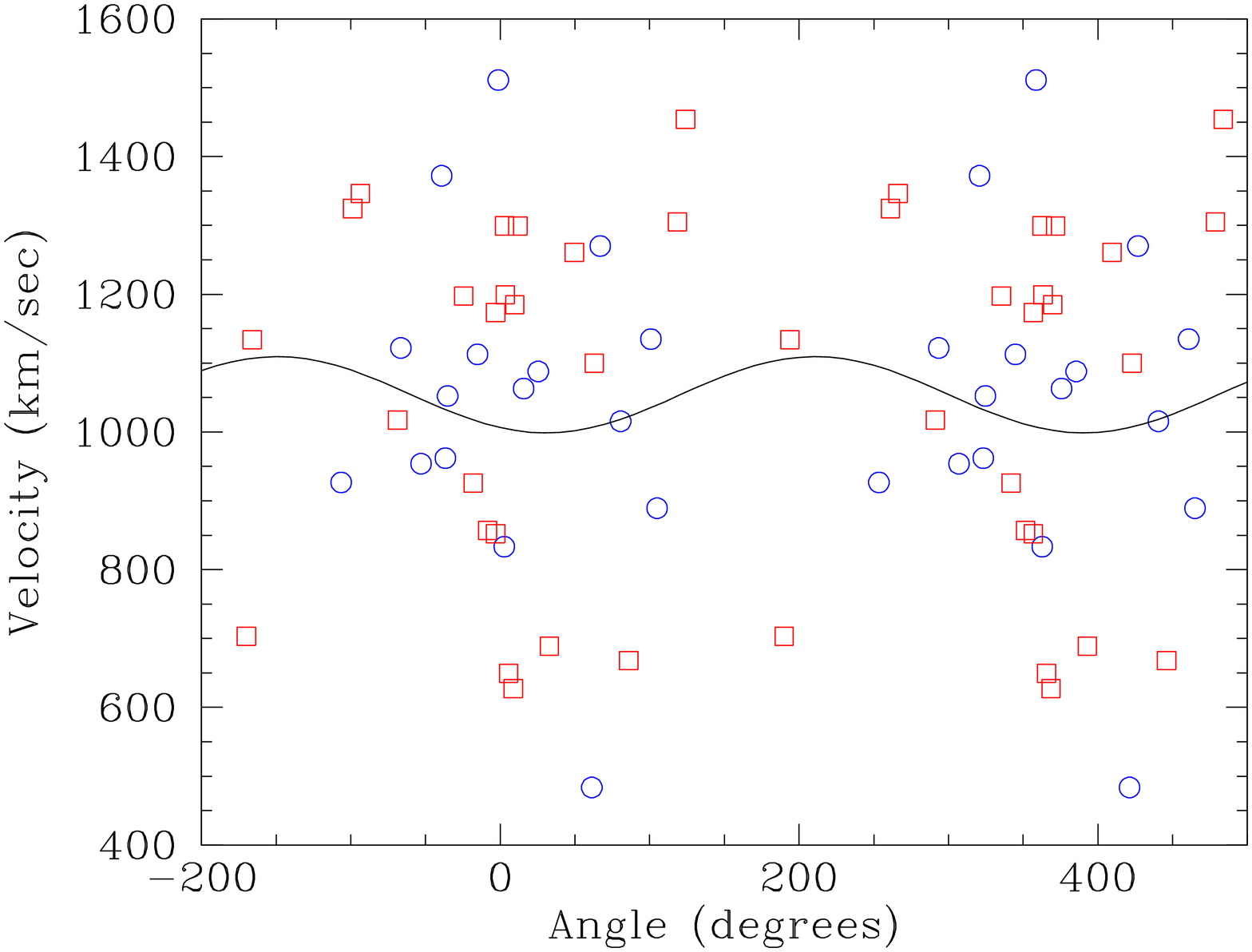,width=15cm,angle=270}}
\caption{Velocity vs azimuthal angle for all (38) NGC 4649 GCs. Blue
GCs are the open circles, and the red GCs are the open squares. The
solid line is the best-fit rotation curve for all GCs, returned by the fitting
code. No significant rotation is seen in any NGC 4649 GC sample.}
\label{rot_all}
\end{figure*}

We have fit the above equation for all three GC samples (total, blue,
and red). Because of the small numbers of GCs, we fixed V$_0$ at the
mean GC velocity (1066 km/sec). The GC mean velocity and systemic
velocity of NGC 4649 differ by about 50 km/s. Since we are measuring
rotation and because our GC sample is not centered on the galaxy, by
using the GC mean rotation, we may be removing rotation at the level
of 50 km/s. Thus, we will likely be  
underestimating the true rotation. We have also looked for
rotation using the systemic velocity, but find small differences. In
either case, rotation is not significant. Figure~\ref{rot_all} shows
the GC velocities as a function of azimuthal angle, with the
``best-fit'' rotation curve for all GCs superimposed.
It is clear that the data in Figure \ref{rot_all} show no rotation
signature. 
This is confirmed by our quantitative fitting, which yields a rotation
of 57 $\pm$ 58 km/s for all GCs, and similar null results for 
blue and red GCs considered separately.
In summary, there is no significant rotation in any of the GC samples.

We have used Monte Carlo simulations to set an upper limit on the
rotation velocity for all GCs. We generate artificial samples with the same
position angle distributions and velocity dispersions as the data, but
with a rotation curve of given amplitude imposed.  The 95\% confidence
upper limit is defined as the rotation velocity for which only 5\% of
these simulations give a rotation velocity as small as that observed
(see Zepf et al. 2000). We find a 95\% upper limit of 150 km/s for the
rotation amplitude of all GCs. Combined with the velocity dispersion
of 256 km/s (Section \ref{dispersion}), we derive an upper limit for
the ratio of rotation amplitude to velocity dispersion of 
v/$\sigma$ $<$ 0.6.

It is of course possible that rotation is present in the NGC 4649 GC
system, but we have failed to detect it. 
The main factor is our poor spatial
and azimuthal coverage (see Figure \ref{rot_all} for our uneven
azimuthal coverage).
As well, while our
total sample of 38 GCs is reasonable, the blue and red GC sample sizes
are smaller than we would like. We plan to acquire more velocities in
our other two fields, which will alleviate the above difficulties and 
allow us to set tighter constraints on the presence or absence of rotation.

\subsection{Dynamical Models}
\label{mlratio}

We use two approaches to measure the M/L profile in NGC 4649. As a
first step, we use spherical, isotropic models applied to the stellar
kinematics and the GC kinematics.  With only 38 GC
velocities, it is hard to go much beyond the isotropic assumption
without additional constraints. While we fully understand the
limitations with these assumptions, we argue that for comparative value,
isotropic models are informative since many previous studies have
relied on these. Our second approach uses fully-general axisymmetric
models. These models are based on Schwarzchild (1979), where one runs a
set of orbits in a specified potential and fits for the orbital
weights to best match the observational contraints. However, since
these models are so general (i.e., allowing anisotropy as a free
parameter spatially in both radius and angle), 38 velocities will be
inadequate to overcome the uncertainties in both the potential and
orbital structure. Fortunately, NGC 4649 has a detailed mass model from
{\it Chandra} X-ray observations (Humphrey et al. 2006). Thus, we use the
M/L profile as determined from X-rays as the input into the orbit-based
models, which then provide the best-fit orbital structure. Both techniques
give the same answer, that NGC 4649 requires a substantial increase in
the M/L at large radii.

\subsubsection{Isotropic Models}

The isotropic models are straightforward and follow the approach
outlined in Pierce et al. (2006a).  Briefly, we use non-parametric
models assuming isotropy (see Gebhardt \& Fischer 1995). We use both
the GC velocities and stellar kinematic data for the inner 70\arcsec\
from Pinkney et al.  (2003).  For the surface brightness profile, we
use the stellar light profiles of Lauer et al. (2005) and Caon et
al. (1990), where the latter extends out to over 700\arcsec. The Lauer
et al. profile is in the $V$-band, and we transform the Caon et
al. $B$-band profile by matching in the overlap region. We find
$B-V=0.98$. We apply an extinction correction of $A_V=0.088$ (Schegel
et al. 1998).
However, since we are fitting the GCs, we should in fact be using
their number density profile. Forbes et al. (2004) compare the GC
profile to the stellar profile. They find that the red GCs are very
similar to the stellar profile, whereas the blue GCs are slightly
flatter. Given the uncertainties in these profiles, and that there are
more velocities from the red GCs, use of the stellar light profile is
reasonable.

\begin{figure*}
\centerline{\psfig{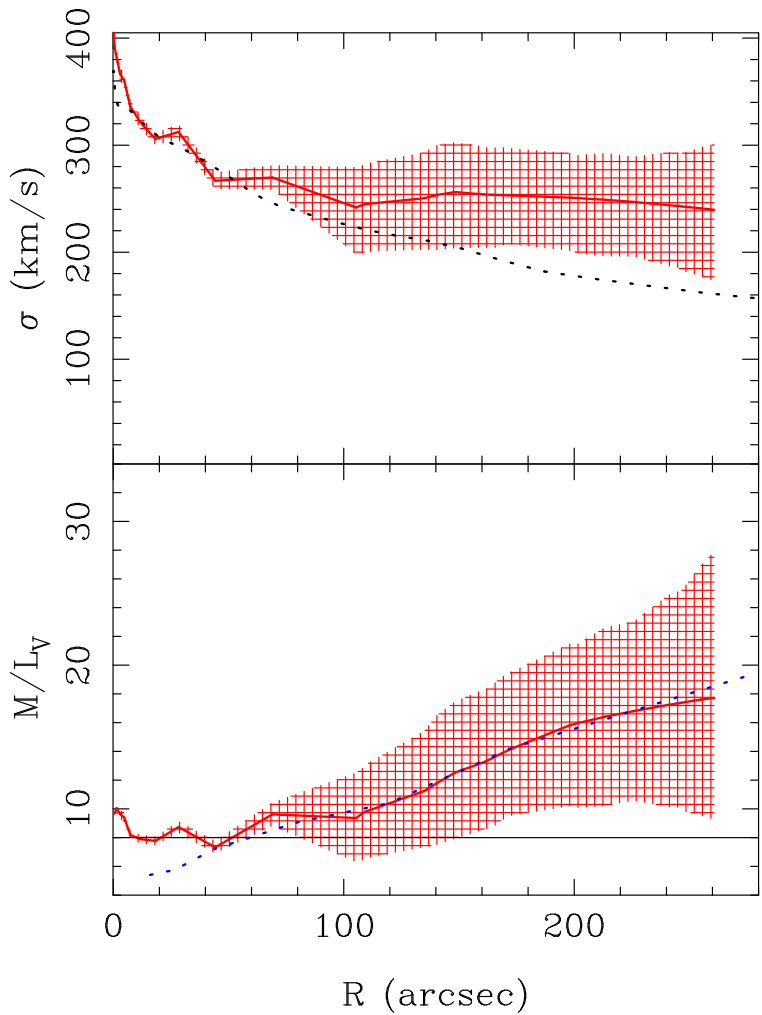}}
\caption{
{\bf Upper Panel:} Velocity dispersion against projected radius
for NGC 4649 GCs and stellar data. The solid (red) line is
the GC data, with 1$\sigma$ errors (red hatched region). The
dashed (black) line is the expected isotropic dispersion profile for a 
constant M/L ratio model.
{\bf Lower Panel:} M/L$_V$ as a function of projected radius for the
NGC 4649 stars, GCs, and X-ray emission. The solid (red) line is the 
profile for the stars and GCs, and the red hatched region are the
1$\sigma$ errors. The dashed (blue) line is the profile for the X-ray
emission, obtained from the X-ray mass profile of Humphrey et al. (2006)
and our V-band light profile.
The horizontal line with M/L$_V$ = 8 represents the value
adopted for the constant M/L ratio model in Section \ref{orbit}.
For our assumed distance of 17.3 Mpc, 100\arcsec\ corresponds
to a linear radius of 8.1 kpc. The effective radius R$_e$ for NGC 4649
varies between 42\arcsec\ in the K-band (Jarrett et al. 2003) to
$\sim$ 80\arcsec\ in the optical (Faber et al. 1989; van der Marel 1991).
}
\label{mlprof}
\end{figure*}

Figure \ref{mlprof} shows the results. The upper panel shows the
combined stellar and GC velocity dispersion (solid (red) line), with 
1$\sigma$ errors (red hatched region). The dashed (black) curve in the upper
panel is the expected isotropic velocity dispersion profile for
a constant M/L ratio (ie. no dark matter). The GC data are 
inconsistent with the constant M/L ratio model over most of the radial
range. The lower panel shows the projected (M/L)$_V$ ratio for the stellar
and GC data with 1 $\sigma$ errors (solid (red) line and hatched region), 
and the dashed (blue) line
is the {\it Chandra} X-ray (M/L)$_V$ profile from Humphrey et al. (2006).
The X-ray analysis of Humphrey et al. provides the total
mass, and we convert to M/L by dividing by our V-band light profile. 
The GC data show an increasing M/L ratio with radius, in
other words implying a DM halo in NGC 4649.
There is excellent agreement between the GC and X-ray mass profiles,
reinforcing the evidence for this DM halo.

\subsubsection{Orbit-Based Models}
\label{orbit}

All aspects of the axisymmetric orbit-based models are provided in
four papers: Gebhardt et al. (2000, 2003) and Thomas et al. (2004,
2005). Thomas et al. (2005) provides the most direct comparison since
it studies dark halo properties. Briefly, we first input a potential,
either from stellar light only or a known mass profile (i.e., the
X-ray potential). We then run a set of orbits in that potential that
covers the available phase space. Thomas et al. (2004) provide a
detailed description of the orbit sampling. For NGC 4649, we use 
approximately $10^4$ orbits and 255 observational constraints. 
We then determine the
non-negative orbital weights that best match the observational
constraints. We use entropy as a way to facilitate the $\chi^2$
minimization over this large set of parameters. The use of entropy as
a way to regularize the models has very little effect on the results
(as we discuss below).

For the potential, we use both the stellar light profile and the X-ray
profile.  The stellar light profile is the same as used above for the
isotropic models, except we now include the proper axis ratio of
0.88. We assume that the galaxy is edge-on. 
We run two sets of orbits: one for a constant stellar M/L
(taking M/L$_V$ = 8; see Figure \ref{mlprof}), and
one with a profile that uses a constant stellar M/L 
(again M/L$_V$ = 8) out to 45\arcsec\
and following the X-ray M/L profile beyond that. At the outer limit of the
model, the increase in the M/L is a factor of 2.5 from the central
regions (inside 45\arcsec). For completeness, we include a central
black hole as reported in Gebhardt et al. (2003); this will likely
effect the kinematics near the center only, but since we desire the
orbital structure throughout the galaxy, it is important to include as
realistic a potential as possible. The model is the same as in Gebhardt
et al. (2003), where we divide the model in spatial bins consisting of
20 radial and 5 angular bins. The orbit sampling as outlined in Thomas
et al. (2005) allows for very efficient coverage of the available
phase-space of the distribution function.

The kinematics come from published results in Pinkney et al. (2003)
and the GC velocities presented here. Pinkney et al. have fairly
extensive stellar kinematics along 3 position angles and go out to
70\arcsec\ along the major axis. For the GCs, we only use those beyond
the stellar data, and we only use them in three radial bins. We
measure the velocity dispersions from the GCs using a
maximum-likelihood approach that includes the uncertainties directly
(as discussed in Pryor \& Meylan 1993).  The three dispersion values
give approximately the same results as the lowess estimator. The three
radial bins are at $R=118, 183, 240$\arcsec\ with dispersion values of
230 $\pm$ 60, 270 $\pm$ 75, 220 $\pm$ 70 km/s, respectively.  Since
the GCs are spread over the full angular range of the galaxy, we sum
up the model angular bins in the same way (as opposed to the stellar
kinematics, which are along particular position angles).

There are two main ways in which to use the orbit-based models. First,
one can use the fit in a $\chi^2$ sense to determine which potential
gives the best fit to the data. In this way, one would run orbits in a
large set of potentials and then choose which one has the lowest
$\chi^2$ as the best model. The second result from the orbit models is
that for a given potential, the model produces a distribution function
that optimizes the fit to the observations. Thomas et al. (2004,
2005), as well as Cappellari et al. (2005), have done many tests
determining the ability to recover the distribution function using
analytic models. If one has adequate radial and angular coverage, then
the distribution function can be recovered robustly. In the case of
NGC 4649, we are unlikely to get detailed information on the
distribution function, but should be able to obtain reliable projections of
it. For example, as shown in Gebhardt et al. (2003) and Gebhardt
(2004), the simple ratio of the radial to tangential dispersions as a
function of radius and angle is a useful and robust quantity. Thus,
for NGC 4649 we concentrate on this ratio. The program obviously
provides both angular dispersions: the azimuthal (parallel to the
equatorial plane) and the polar (perpendicular to the equatorial
plane). We add in quadrature the two angular dispersions and average,
and compare this quantity to the radial dispersion. Thus, isotropic 
models have radial and tangential dispersions equal to each other.

\begin{figure*}
\centerline{\psfig{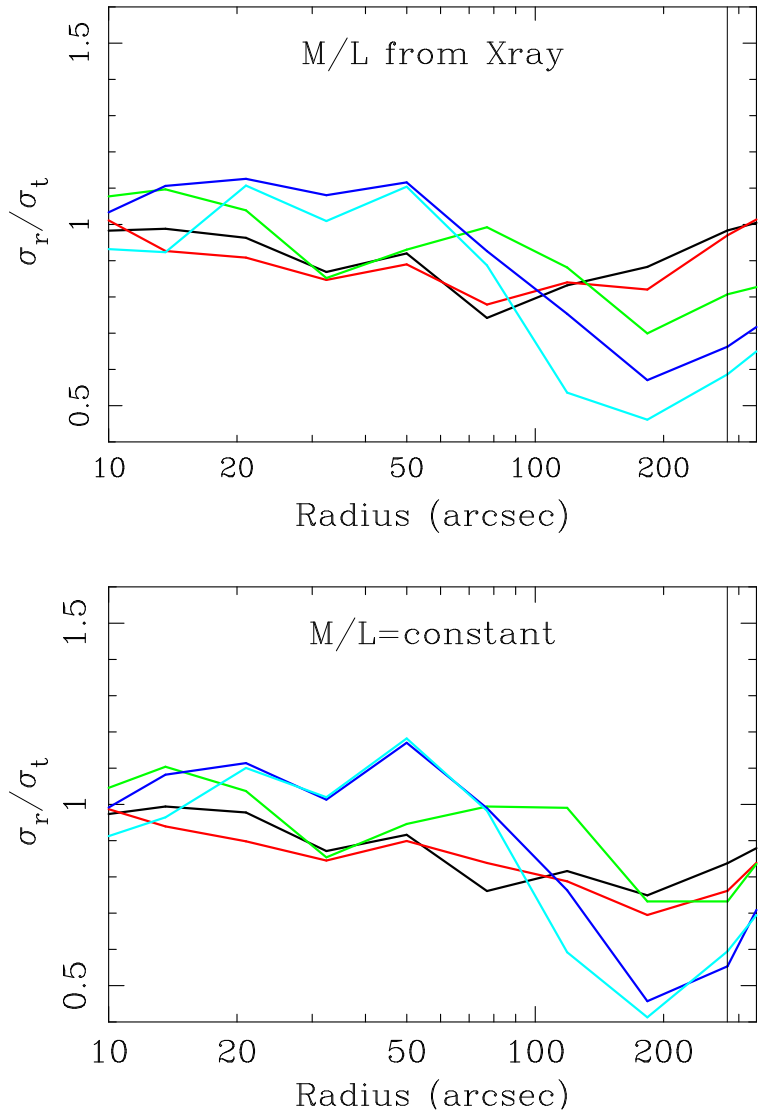}}
\caption{The ratio of the radial to
the tangential velocity dispersion along the five position angles
in the orbit-based models. A ratio of one would imply an isotropic
model. The vertical solid line is the radial extent of the
observations; results beyond that radii are not meaningful.
The top panel assumes a mass profile as determined from the X-ray
observations of Humphrey et al., and the bottom panel assumes
a mass profile from the stellar light with a constant M/L. Using
the mass profile from the X-rays produces less tangential bias
at large radii, although some is still present implying either
a need for a larger M/L increase or a real effect in the globular
cluster orbits.
}
\label{mlorbit}
\end{figure*}

Figure ~\ref{mlorbit} shows the internal dispersion ratio for the 
potentials using the M/L profile from X-rays (upper panel) and a 
constant M/L (lower panel), along each position angle. 
For the X-ray profile,
the orbital structure becomes slightly tangentially biased beyond
100\arcsec; the average ratio is about 0.8. For the constant M/L
model, the average ratio is about 0.7. The stronger tangential bias
in the constant M/L model is 
what one would expect if a dark halo is present but not included in
the dynamical models. One way to increase the projected dispersions at
large radii is to put more energy into the tangential component.

Due to the large uncertainties in the kinematics from the GCs, we
have not explored other dark halo models beyond the X-ray potential.
However, we note that the overall fit to the data is better with the
X-ray potential than with the constant M/L model. The change in
$\chi^2$ is 5, implying a signficance of greater than 95\%. Thus, with
these data and the X-ray potential alone, we find a strong significance
for a dark halo. A further test would be to try a variety of dark halo
models. Since we only tried one, it is very likely that the minimum
$\chi^2$ will be even lower, implying a yet higher
significance. Thus, we have not determined the best fit dark halo
model for this dataset; this implies that we have not found the best
fit orbital structure as well. Presumably, if the dark halo were more
massive than the X-ray model suggests, we would find an orbital
structure that is more consistent with isotropy. However, this test
does demonstrate the power of combining X-ray data and kinematics. If we
take the X-ray potential as truth, then the underlying orbital distribution
is as given in Figure ~\ref{mlorbit}; i.e., the GCs are nearly isotropic
out to 100\arcsec, and tangentially biased beyond that radius.

\section{Discussion}
\label{discussion}

We have not detected any rotation in the NGC 4649 GC system, with 
v/$\sigma$ $<$ 0.6 for all GCs. As discussed in the Introduction, there
is considerable variation in the rotation properties of GC systems
amongst early-type galaxies.
Projection effects certainly play some role in explaining 
this variation, 
but these effects have not been properly studied to date.
A range of rotation properties is also seen from 
PNe data, with several ellipticals showing little rotation 
(v/$\sigma$ $\sim$ 0.2; Romanowsky 2004). Yet most simulations of 
elliptical galaxy formation predict significant amounts of angular momentum
in their outer halos (e.g. Barnes 1992; Vivitska et al. 2002).
One possibility is that angular momentum is transported beyond
the radial range of our data, which is $\sim$ 20 kpc or 3.2$-$6.5 R$_e$.
Simulations also show that the angular momentum of DM
halos is generally larger for major mergers, and lower for multiple
accretion of satellites (Gerhard 2005). Thus, the merging history of 
ellipticals may also play a role in determining their rotation. It is
interesting to note in this context that there is no evidence for a
a major merger (or indeed any recent interaction) in NGC 4649. 
Pierce et al. (2006b) have found several young (2-3 Gyr) 
GCs in NGC 4649. The origin of these young GCs is not clear, but 
the fact that GCs younger and older than 9 Gyr have very similar velocity
dispersions (see Section \ref{dispersion}) argues against the young 
GCs having been formed in a merger.

We have also been able to put constraints on the orbital structure of
the GCs, through the use of orbit-based models with a specified
potential. We find that the GC orbits are close to isotropic out to
100\arcsec, but become tangentially biased at larger radii
(Figure \ref{mlorbit}). Unfortunately, there are very few early-type
galaxies with which we can compare our results.  An analysis similar
to ours for NGC 4649 has been carried out for M87 (Cote et al. 2001)
and M49 (Cote et al. 2003). In both galaxies, the entire GC system is
close to isotropic at all radii. There are differences, however, when
the metal-poor and metal-rich GCs are considered separately. In M87,
the metal-poor GCs are tangentially biased at all radii, particularly
within 15 kpc, while the metal-rich GCs are isotropic or slightly
tangentially biased within 15 kpc, and radially biased beyond this
radius. In M49, the metal-poor GCs are close to isotropic, while the
metal-rich GCs become tangentially biased at large radius.  Richtler
et al. (2004) were able to place weak constraints on the orbital
structure of the NGC 1399 GCs, finding that the metal-rich GCs are
close to isotropic, and that the metal-poor GCs have a slight
tangential bias.  Overall then, existing data show that GCs in
early-type galaxies generally have orbits that are close to isotropic,
but there may be differences in the orbital structure of metal-poor
and metal-rich GC subsystems. Unfortunately, our small sample size
does not allow us to investigate such possible differences in the
metal-poor and metal-rich GCs in NGC 4649. The models presented in
this paper use axiysmmetric orbit-based models, which are the most
general models one can run assuming axisymmetry, and thus should
provide the most fair description of the orbital structure. The
previous models listed above are all based on spherical symmetry,
and it may be that once more realistic
models are used, a clear picture of the outer structure may emerge.

We have found evidence for a DM halo in NGC 4649
from both spherical, isotropic models
and axisymmetric, orbit-based models. The isotropic models give a
M/L$_V$ of 9$-$28 at 260\arcsec\ radius ($\sim$ 21 kpc; Figure \ref{mlprof}).
The isotropic GC M/L profile (Figure \ref{mlprof}) gives excellent
agreement with the {\it Chandra} M/L profile (Humphrey et al. 2006).
de Bruyne et al. (2001) also found evidence for a 
DM halo from their stellar kinematic data extending to $\sim$ 90\arcsec.
Figure \ref{mlprof} shows that the NGC 4649 M/L ratio is $\sim$ 8 near the
galaxy center, and starts increasing at $\sim$ 50\arcsec. 
There is a significant difference between the effective radius 
R$_{e}$ for NGC 4649 in the optical and IR, with optical values between
70$-$80\arcsec\ (RC3; Faber et al. 1989; Van der Marel 1991), and a
2MASS K-band value of 42\arcsec\ (Jarrett et al. 2003). In any case,
both X-ray and GC data suggest that NGC 4649 is becoming
DM-dominated beyond $\sim$ 1 R$_e$, as has been found for other ellipticals
(Gerhard 2005; Romanowsky 2005; Mamon \& Lokas 2005). 

DM halos are commonly found in other early-type galaxies from GC,
X-ray, and stellar kinematic data (see Introduction; Gerhard 2005;
Napolitano et al. 2005). However, PNe data have been used to argue
against the existence of DM halos in early-type galaxies
(e.g. Romanowsky et al. 2003). As discussed in the Introduction, this
may be because PNe in early-type galaxies are on preferentially radial
orbits.  For cluster galaxies there is the added complication of
potential contamination from intra-cluster PNe (e.g. Feldmeier et
al. 2004).  Unfortunately, there are no published PNe data for NGC
4649 to compare our GC results with. 
GC spectroscopy allows both the
determination of velocities and ages/abundances, allowing us to
investigate whether kinematics depends on age. However, even with
8m-class telescopes, it is difficult to obtain the high S/N spectra
needed for reliable age determinations, so there is little age data
available at present.  We have found that exclusion of four young
(2$-$3 Gyr: \#89, 175, 502, 1443) and two intermediate-age (5$-$6 Gyr:
\#517, 1182) GCs in NGC 4649 (see Pierce et al. 2006b) has no
significant effect on the M/L profile, but our GC sample size is not
large. More kinematic data for a range of tracers, including PNe, GC,
stellar, and X-ray, are needed for NGC 4649 and other early-type
galaxies.

\section{Conclusions}
\label{conclusions}

We have used Gemini/GMOS to obtain
spectra for 38 confirmed globular clusters in NGC 4649. The recession
velocities of these clusters have been used to study the cluster
kinematics and dark matter content of NGC 4649. Our main results are:

\begin{itemize}

\item The GC velocity dispersion profile is
constant with radius, for all GCs, and also for the blue and red GCs
separately.

\item We detect no rotation in the GC system, both for
all clusters, and for blue and red samples separately. We are able to
place upper limits on the ratio of rotation velocity to velocity dispersion
v/$\sigma$ $<$ 0.6 for all GCs.
Further data to improve our spatial and azimuthal coverage,
and our sample size, are needed to make further progress.  

\item Both spherical, isotropic and axisymmetric, orbit-based 
dynamical models
strongly support the presence of a dark matter halo in NGC 4649. For
the isotropic models there is excellent agreement between the
stellar plus GC M/L profile and the X-ray M/L profile from
Humphrey et al. (2006).

\item Within $\sim$ 100\arcsec\ radius, the GC orbits are close to isotropic,
while at larger radius the orbits become tangentially biased.

\end{itemize}

\section*{Acknowledgments}

We thank the anonymous referee for a thorough reading of this paper,
and for several useful suggestions which have significantly improved
the paper.
We'd like to thank Inger Jorgensen and Gemini staff astronomers
for wonderful support for our GMOS program. We are very grateful to
Soeren Larsen for sharing his HST data with us. DF thanks the ARC
for its financial support. SEZ acknowledges support for this work
in part from the NSF grant AST-0406891 and from the Michigan State
University Foundation. This research was supported in part by a 
Discovery Grant awarded to DAH by the Natural Sciences and Engineering
Research Council of Canada (NSERC). KG acknowledges support from
NSF CAREER grant AST-0349095. JCF and
FF have been supported with grants from CONICET and Agencia Nac. de
Promocion Cientifica, Argentina.
These data were based on observations obtained at the Gemini
Observatory, which is operated by the Association of Universities for
Research in Astronomy, Inc., under a cooperative agreement with the
NSF on behalf of the Gemini partnership: the National Science
Foundation (United States), the Particle Physics and Astronomy
Research Council (United Kingdom), the National Research Council
(Canada), CONICYT (Chile), the Australian Research Council
(Australia), CNPq (Brazil) and CONICET (Argentina). The Gemini
program ID is GN-2002A-Q13. This research has made use of the
NASA/IPAC Extragalactic Database (NED), which is operated by the
Jet Propulsion Laboratory, Caltech, under contract with the 
National Aeronautics and Space Administration.

\label{lastpage}

\end{document}